\documentclass{PoS}

\usepackage{mathrsfs,bbold}
\usepackage{amsmath,amssymb,graphicx,listings,xcolor,tikz} 
\usepackage{epstopdf,cancel}
\usepackage{framed, color}

\newcommand*{\mathcolor}{}
\def\mathcolor#1#{\mathcoloraux{#1}}

\newcommand*{\mathcoloraux}[3]{%
  \protect\leavevmode
  \begingroup
    \color#1{#2}#3%
  \endgroup
}

\newcommand{\eps}{\varepsilon}
\newcommand{\bea}{\begin{eqnarray}}
\newcommand{\eea}{\end{eqnarray}}

\newcommand{\be}{\begin{equation}} 
\newcommand{\ee}{\end{equation}}
\newcommand{\ba}{\begin{array}}
\newcommand{\ea}{\end{array}}

\title{Searches for Sterile Neutrinos at Future Electron-Proton Colliders}

\ShortTitle{Searches for Sterile Neutrinos at Future Electron-Proton Colliders}

\author{\speaker{Oliver Fischer}\\
        \normalsize	Department of Physics, University of Basel, \\ 
	\normalsize 	Klingelbergstr.\ 82, CH-4056 Basel, Switzerland.\\
        E-mail: \email{oliver.fischer@unibas.ch}}

\author{Stefan Antusch\\
	\normalsize	Department of Physics, University of Basel, \\ 
	\normalsize 	Klingelbergstr.\ 82, CH-4056 Basel, Switzerland;\\ 
	
	\vspace{-12pt}
	\normalsize	Max-Planck-Institut f\"ur Physik (Werner-Heisenberg-Institut), \\ 
	\normalsize 	F\"ohringer Ring 6, D-80805 M\"unchen, Germany.\\
        E-mail: \email{stefan.antusch@unibas.ch}}

\abstract{Sterile neutrinos are an attractive extension of the Standard Model of elementary particles towards including a mechanism for generating the observed light neutrino masses. We discuss that when an approximate protective ``lepton number''-like symmetry is present, the sterile neutrinos can have masses around the electroweak scale and potentially large neutrino Yukawa couplings, which makes them well testable at planned future particle colliders. We systematically discuss the production and decay channels for sterile neutrinos at electron-proton colliders and give a complete list of the leading order signatures for sterile neutrino searches. We highlight several novel search channels and present a first look at the possible sensitivities for the active-sterile mixing parameters and the heavy neutrino masses. We also compare the performance of electron-proton colliders with the ones of proton-proton and electron-positron colliders, and discuss the complementarity of the different collider types.}

\FullConference{XXV International Workshop on Deep-Inelastic Scattering and Related Subjects\\
		3-7 April 2017\\
		University of Birmingham, UK}

\begin{document}

\section{Introduction}
\begin{figure}
\begin{center}
\includegraphics[scale=0.25]{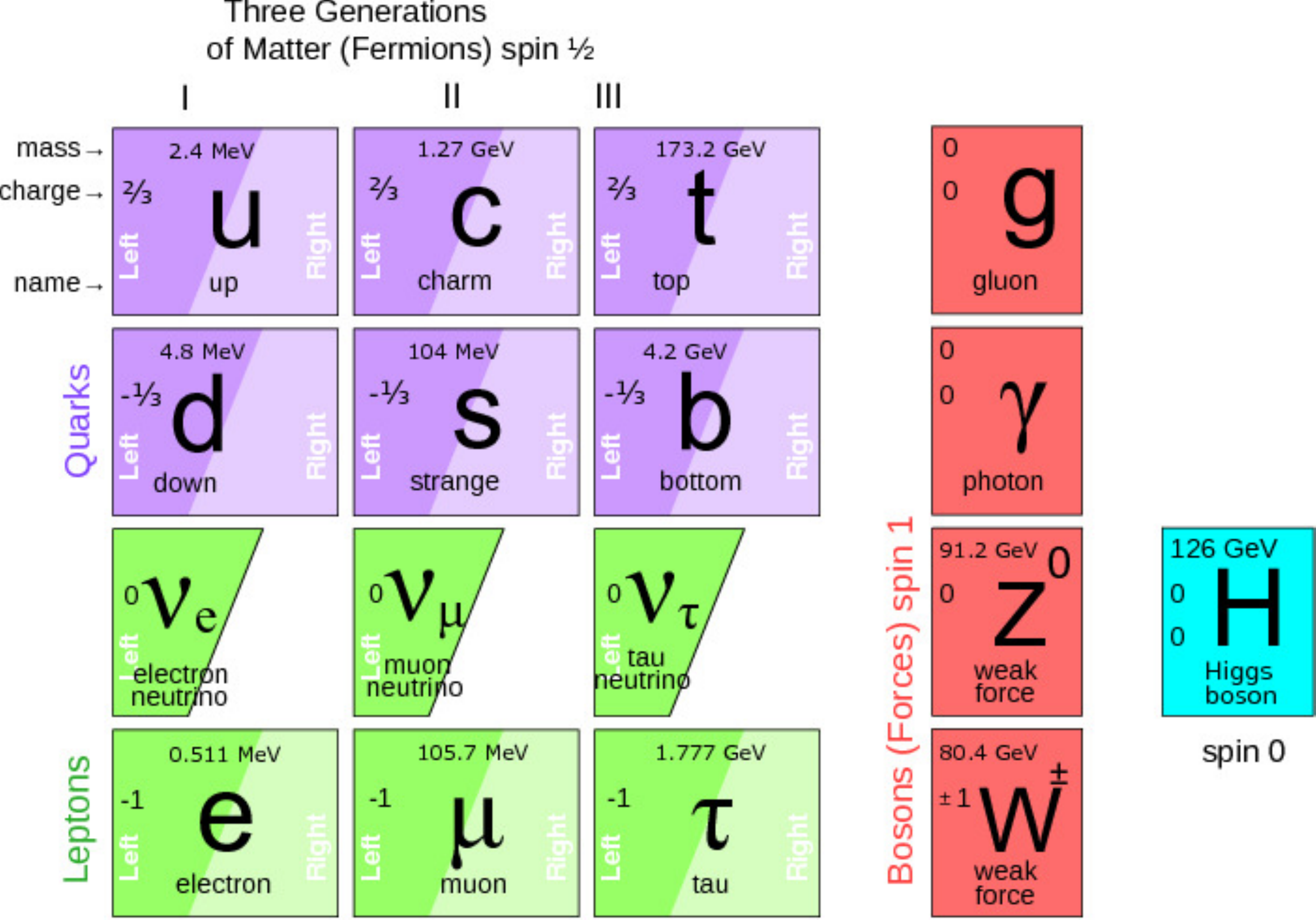}
\end{center}
\caption{Particle content of the SM \cite{MarcoDrewes}, including the quarks and leptons, the gauge bosons, and the Higgs boson. The missing right-chiral counterparts of the neutrinos is indicated pictorially.}
\label{fig:SM}
\end{figure}
\noindent
The observation of neutrino oscillations requires two mass splittings between the three active neutrinos, which in turn implies that at least two of the active neutrino species must have non-zero masses.
Since the Standard Model (SM) does not contain neutrino masses, neutrino oscillations are evidence for physics beyond the SM (BSM).

The neutrino masses can easily be realised with Yukawa interactions, when the SM field content, cf.\ fig.~\ref{fig:SM} is ``completed'' with a number of right-handed neutrinos \cite{Minkowski:1977sc,Mohapatra:1979ia,Yanagida:1980xy,GellMann:1980vs}. Such new particles would be completely neutral with respect to the SM gauge group, which is why we will refer to them as {\it sterile} neutrinos from now on.
Their lack of quantum numbers also allows for mass terms that mix up only the sterile neutrinos, such that the mass matrix for the neutral fermions contains the Dirac-type masses that emerge after breaking of the electroweak symmetry as well as the Majorana-like masses of the sterile neutrinos.

New physics models with sterile neutrinos have very attractive features: they can address the dark matter (DM) problem in minimal models, for instance the $\nu$MSM (see e.g. ref.~\cite{Canetti:2012vf,Canetti:2012kh}), or in slightly expanded versions (see e.g.\ ref.~\cite{Heurtier:2016iac}); they allow for leptogenesis at the GUT scale (see e.g.\ refs.~\cite{Davidson:2002qv,Hambye:2003rt}), the TeV scale (see e.g.\ refs.~\cite{Pilaftsis:1997jf,Pilaftsis:2003gt,Dev:2014laa,Hambye:2016sby}), and even for low masses \cite{Drewes:2012ma,Drewes:2016gmt,Drewes:2016jae}; they account for the light neutrinos' masses and mixings.
See e.g.\ ref.~\cite{Drewes:2013gca} for a review on the phenomenology of sterile neutrinos in the early universe, and ref.~\cite{Antusch:2016ejd} for an overview of searches for sterile neutrinos at particle colliders.

\section{The type I seesaw mechanism}
\noindent
When a number of right-handed or sterile neutrinos are added to the field content of the SM, the diagonalisation of the mass matrix for the active and the sterile neutrinos yields a number of mass eigenstates. 
This is commonly referred to as the type I seesaw mechanism.

In order to understand how it works, we consider the ``na\"ive'' version of this mechanism, with exactly one active and one sterile neutrino, with a Dirac mass $m_D = |y_\nu| v_{\rm EW}$, where $y_\nu$ and $v_{\rm EW}$ are the neutrino Yukawa coupling and the Higgs vacuum expectation value, respectively, and the sterile neutrino mass $M_R$.
In the case where $m_D \ll M_R$, the light neutrino mass can be approximated by 
\begin{equation}
m_\nu = {1 \over 2} \frac{v_\mathrm{EW}^2 |y_\nu|^2}{ M_R}\,.
\label{eq:naiveseesaw}
\end{equation}
The upper bounds on the light neutrino mass scale can be naturally explained when $M_R\sim M_{\rm GUT}$, or, alternatively, for tiny values of $|y_\nu|$. 

We can expand this simplistic picture of the seesaw mechanism to two active and sterile neutrinos. In this case, the Yukawa and Sterile neutrino masses become $2\times2$ matrices, for instance
\begin{equation}
Y_\nu = \begin{pmatrix} {\cal O}(y_\nu) & 0 \\  0 & {\cal O}(y_\nu) \end{pmatrix},\qquad M_N = \begin{pmatrix} M_R & 0 \\ 0 & M_R (1+ {\varepsilon}) \end{pmatrix}\,,
\end{equation}
where the parameter ${\varepsilon}$ suggestively indicates the possibility of breaking the mass degeneracy first and the second ``family'' of light and heavy neutrino mass eigenstates:
\begin{equation}
m_{\nu_i} \approx \frac{v_\mathrm{EW}^2 {\cal O}(y_\nu^2)}{M_R}(1- \delta_{i2}{\varepsilon})\,.
\end{equation} 
Although this scenario allows to explain the observation of the oscillations, also here the upper bounds on the light neutrino mass scale require very large $M_R$ or very small $y_\nu$.

It is possible to realize a type I seesaw mechanism with large neutrino Yukawa couplings and sterile neutrino masses around the electroweak scale when specific structures of the mass matrix are introduced, see for instance ref.~\cite{Gavela:2009cd} and references therein.
This can be achieved, for instance, with additional symmetries and does not necessarily require any tuning of parameters. A simple example for such a seesaw scenario with two active and two sterile neutrinos with an additional symmetry is for instance given by
\begin{equation}
Y_\nu = \begin{pmatrix} {\cal O}(y_\nu) & 0 \\  {\cal O}(y_\nu) & 0  \end{pmatrix},\qquad \begin{pmatrix} 0 & M_R  \\ M_R & {\varepsilon} \end{pmatrix}\,,
\label{eq:massmatrix}
\end{equation}
with the small parameter ${\varepsilon}$ breaking the symmetry. In this scenario a mass for one of the light neutrinos is generated,
\begin{equation}
m_{\nu_i} \approx 0 + {\varepsilon} \frac{v_\mathrm{EW}^2 {\cal O}(y_\nu^2)}{M_R^2}\,,
\end{equation}
that is, the masses are zero in the symmetric limit, and a small perturbation generates a mass proportional to ${\varepsilon}$, the amount of symmetry breaking.
The bounds on the light neutrino mass scale do not restrict the combination of the neutrino Yukawa coupling and the sterile neutrino mass.
This implies, that large neutrino mixings and heavy neutrinos on the electroweak scale can be compatible with the observed smallness of the light neutrino masses.\footnote{Note that alternatively one can also consider a small perturbation of the structure in eq.~(\ref{eq:massmatrix}) in the zero elements of $Y_\nu$, and again the generated light neutrino mass(es) are then proportional to this symmetry breaking parameter.}

\begin{figure}
\begin{center}
\includegraphics[width=0.7\textwidth]{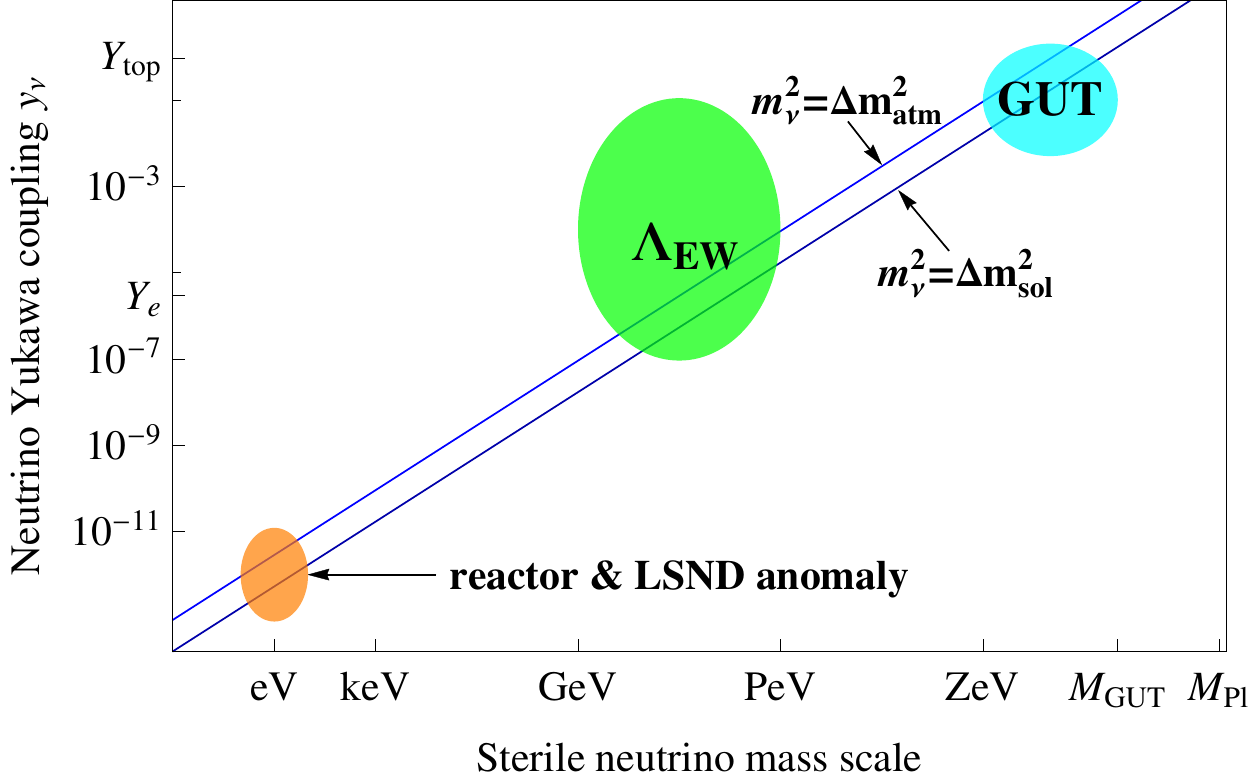}
\end{center}
\caption{Parameter space of the type I seesaw scenario. The diagonal lines denote the na\"ive type I seesaw relation from eq.\ \eqref{eq:naiveseesaw}, where the light neutrino mass is taken to be the atmospheric and the solar mass difference, respectively. The green area denotes the parameter space that may be testable at colliders, the blue area the realm of GUT motivated high scale seesaw scenarios, and the orange area indicates where the reactor and LSND anomalies are situated.}
\label{fig:bigpicture}
\end{figure}

We show the parameter space of the type I seesaw scenario in fig.~\ref{fig:bigpicture}. Therein, the modulus of the neutrino Yukawa coupling is shown on the y-axis, and the x axis denotes the sterile neutrino mass scale. The na\"ive type I seesaw relation from eq.\ \eqref{eq:naiveseesaw} is shown by the two diagonal lines for $m_{\nu_i}$ being the atmospheric and the solar mixing, respectively. Parameter combinations above these lines require either fine tuning or a symmetric seesaw mechanism.
The green area denotes the parameter sets that can be tested with high-energy collision experiments.

\section{The symmetry protected sterile neutrino scenario}
\noindent
As benchmark model we use the Symmetry Protected Seesaw Scenario (the SPSS, see \cite{Antusch:2015mia}), which does capture the relevant features of seesaw models that feature a symmetry which prevents the light neutrino masses from direct contributions of the neutrino-Yukawa couplings. 

The SPSS includes one pair of sterile neutrinos $N_R^I$ $(I=1,2)$ and a protective ``lepton-number-like'' symmetry, under wich $N_R^1$ ($N_R^2)$ carries the same (opposite) charge as the left-handed $SU(2)_L$ doublets $L^\alpha,\,\alpha=e,\mu,\tau$. 
In this model, the masses of the light neutrinos and other lepton-number-violating effects can be introduced together with small symmetry-breaking terms, but are not considered in the following.

\subsection{The model}
\noindent
The  Lagrangian density  of a generic seesaw model with two sterile neutrinos with an exact ``lepton-number-like'' symmetry is given by
\begin{equation}
\mathscr{L} \supset \mathscr{L}_\mathrm{SM} -  \overline{N_R^1} M N^{2\,c}_R - y_{\nu_{\alpha}}\overline{N_{R}^1} \widetilde \phi^\dagger \, L^\alpha+\mathrm{H.c.}\;.
\label{eq:lagrange}
\end{equation}
In this equation $\mathscr{L}_\mathrm{SM}$ contains the known SM fields, $L^\alpha$ and $\phi$ denote the lepton and Higgs doublets, respectively, and the kinetic terms of the sterile neutrinos were omitted. The $y_{\nu_{\alpha}}$ are complex-valued neutrino Yukawa couplings and the sterile neutrino mass parameter $M$ can be chosen real without loss of generality.

In the SPSS, the mass matrix of the two sterile neutrinos and the neutrino Yukawa matrix take the form of eq.\ \eqref{eq:massmatrix}, with $Y_\nu$ being a $2\time 3$ matrix, and we consider the case of the ``lepton-number-like'' symmetry being exact, i.e.\ $\varepsilon=0$.

After EW symmetry breaking, the $5 \times 5$ mass matrix of the electrically neutral leptons emerges, which can diagonalised with the unitary $5 \times 5$ leptonic mixing matrix. Neglecting  ${\cal O}(\theta^2)$ corrections to the heavy neutrinos, the mass eigenvalues are exactly 0 for the three light neutrinos, and $M$ for the two heavy ones.
The mixing of the active and sterile neutrinos can be quantified by the mixing angles, defined as
\begin{equation}
\theta_\alpha = \frac{y_{\nu_\alpha}^{*}}{\sqrt{2}}\frac{v_\mathrm{EW}}{M}\,.
\label{def:thetaa}
\end{equation}
With the leptonic mixing angles the leptonic mixing matrix can be expressed as:
\begin{equation} 
U = \left(\begin{array}{ccccc} 
{\cal N}_{e1}	& {\cal N}_{e2}	& {\cal N}_{e3}	& - \frac{\mathrm{i}}{\sqrt{2}}\, \theta_e & \frac{1}{\sqrt{2}} \theta_e 	\\ 
{\cal N}_{\mu 1}	& {\cal N}_{\mu 2}  	& {\cal N}_{\mu 3}  	& - \frac{\mathrm{i}}{\sqrt{2}}\theta_\mu & \frac{1}{\sqrt{2}} \theta_\mu  \\
{\cal N}_{\tau 1}	& {\cal N}_{\tau 2} 	& {\cal N}_{\tau 3} 	& - \frac{\mathrm{i}}{\sqrt{2}} \theta_\tau & \frac{1}{\sqrt{2}} \theta_\tau \\  
0	   	& 0		& 0	&  \frac{ \mathrm{i}}{\sqrt{2}} & \frac{1}{\sqrt{2}}\\
-\theta^{*}_e	   	& -\theta^{*}_\mu	& -\theta^{*}_\tau &\frac{-\mathrm{i}}{\sqrt{2}}(1-\tfrac{1}{2}\theta^2) & \frac{1}{\sqrt{2}}(1-\tfrac{1}{2}\theta^2)
\end{array}\right)\,.
\label{eq:mixingmatrix}
\end{equation}
The elements of the non-unitary $3\times 3$ submatrix ${\cal N}$, which is the effective mixing matrix of the three active neutrinos, i.e.~the Pontecorvo--Maki--Nakagawa--Sakata (PMNS) matrix, are given as 
\begin{equation}
{\cal N}_{\alpha i} = (\delta_{\alpha \beta} - \tfrac{1}{2} \theta_{\alpha}\theta_{\beta}^*)\,(U_\ell)_{\beta i}\,,
\label{eq:matrixN}
\end{equation}
with $U_\ell$ being a unitary $3 \times 3$ matrix. In the limit of the protective symmetry being exact, the SPSS introduces four parameters (beyond the SM) that are relevant for collider phenomenology: the moduli of the neutrino Yukawa couplings ($|y_{\nu_e}|$, $|y_{\nu_\mu}|$, $|y_{\nu_\tau}|$) and the Majorana mass $M$.

\subsection{Weak interactions of the light and heavy neutrinos}
\noindent
Leptonic mixing gives rise to the light and heavy neutrino mass eigenstates that both interact with the weak gauge bosons. 
This interaction can be expressed with the weak currents in the mass basis,
\begin{subequations}\label{eq:weakcurrentmass}
\begin{eqnarray}\label{eq:chargedweakcurrentmass}
j_\mu^\pm & = & \sum\limits_{i=1}^5 \sum\limits_{\alpha=e,\mu,\tau}\frac{g}{\sqrt{2}} \bar \ell_\alpha\, \gamma_\mu\, P_L\, U_{\alpha i}\, \tilde n_i\, + \text{ H.c.}\,, \\
j_\mu^0 & = & \sum\limits_{i,j=1}^5 \sum\limits_{\alpha=e,\mu,\tau}\frac{g}{2\,c_W} \overline{\tilde n_j}\, U^\dagger_{j\alpha}\, \gamma_\mu\, P_L\, U_{\alpha i}\, \tilde n_i\,, 
\end{eqnarray}
\end{subequations}
where $g$ is the weak coupling constant, $c_W$ is the cosine of the Weinberg angle and $P_L = {1 \over 2}(1-\gamma^5)$ is the left-chiral projection operator, and where the mass eigenstates $\tilde n_j$ of the light and heavy neutrinos are defined via the active and sterile neutrinos:
\begin{equation}
\tilde n_j = \left(\nu_1,\nu_2,\nu_3,N_4,N_5\right)^T_j = U_{j \alpha}^{\dagger} n_\alpha^{}\,, \qquad n = \left(\nu_{e_L},\nu_{\mu_L},\nu_{\tau_L},(N_R^1)^c,(N_R^2)^c\right)^T\,.
\end{equation}
The neutrino mass eigenstates interact with the Higgs boson via the Yukawa interaction in the Lagrangian density in eq.\ \eqref{eq:lagrange}, which, in the mass basis, is given by
\begin{equation}
\frac{\sqrt{2}\,M}{v_\mathrm{EW}} \left[\sum\limits_{i=1}^3 \left(\vartheta_{i4}^* \overline{N_4^c}+ \vartheta_{i5}^*\overline{N^c_5}\right) \phi^0 \nu_i + \sum\limits_{j=4,5} \vartheta_{jj}^* \overline{N^c_j} \phi^0 N_j \right] +\text{ H.c.}
\end{equation}
with $\vartheta_{ij} = \sum_{\alpha=e,\mu,\tau} U^\dagger_{i\alpha}U_{\alpha j}^{}$.

\subsection{Indirect constraints from present precision data}
\noindent
As is discussed in ref.~\cite{Antusch:2014woa}, the presence of sterile neutrinos modifies the theory predictions for the Electroweak Precision Observables. In this reference an updated bound from a global set of precision observables was extracted in the so-called ``Minimal Unitarity Violation'' scheme. 
In addition to the EWPOs, this bound also includes other low energy precision observables:

\begin{itemize}
\item The lepton universality observables, given by the ratios of decay rates: $R^X_{\alpha\beta}=\Gamma^X_\alpha/\Gamma^X_\beta$, where $\Gamma^X_\alpha$ denotes a  decay width including a charged lepton $\ell_\alpha$ and a neutrino. 

\item Measurements of lepton flavour violating processes, in particular from charged leptons: $\ell_\rho\to \ell_\sigma \gamma$. The presently strongest constraint comes from the MEG collaboration\cite{Adam:2013mnn}. 

\item The unitarity of the first row of the CKM matrix. 

\item The weak mixing angle from the NuTeV experiment, using the recent update in ref.\ \cite{Bentz:2009yy}, wherein the tension with other precision data has been removed. 

\item Measurements of parity violation at energies far below the $Z$ boson mass, yieling the weak mixing angle with a precision below the percent level, see e.g.\ the reviews \cite{Erler:2004cx,Kumar:2013yoa} for an overview. 

\end{itemize}
The analysis yields the following highest posterior density (HPD) intervals at 68\% ($1\sigma$) Bayesian confidence level (CL):\cite{Antusch:2014woa}
\begin{equation}
\begin{array}{ccl} \epsilon_{ee}  & = &-0.0012 \pm 0.0006 \\ | \epsilon_{\mu\mu} | & < & 0.00023 \\ \epsilon_{\tau\tau} & = & -0.0025 \pm 0.0017 \end{array} \qquad  \begin{array}{ccl} | \epsilon_{e\mu} | & < & 0.7 \times 10^{-5} \\ | \epsilon_{e\tau} | & < & 0.00135 \\ | \epsilon_{\mu\tau} | & < & 0.00048 \,,\end{array}
\label{MUVresult}
\end{equation}
The best fit points for the off-diagonal $\eps_{\alpha\beta}$ and for $\eps_{\mu\mu}$ are at zero.
We remark that there is a weak statistical preference for non-zero mixing for $\varepsilon_{ee}$ at two sigma confidence.
The non-unitarity parameters can be expressed in terms of the active-sterile mixing parameters: $\varepsilon_{\alpha\alpha} = -\theta_\alpha^*\theta_\alpha^{}$.

The resulting constraints on the mixing parameters are shown in fig.\ \ref{fig:presentconstraints} by the solid, dashed, and dotted lines for $|\theta_e|,\,|\theta_\tau|,$ and $|\theta_\mu|$, respectively.

\subsection{Present constraints from direct searches}
\begin{figure}
\begin{minipage}{0.49\textwidth}
\begin{center}\includegraphics[width=1.0\textwidth]{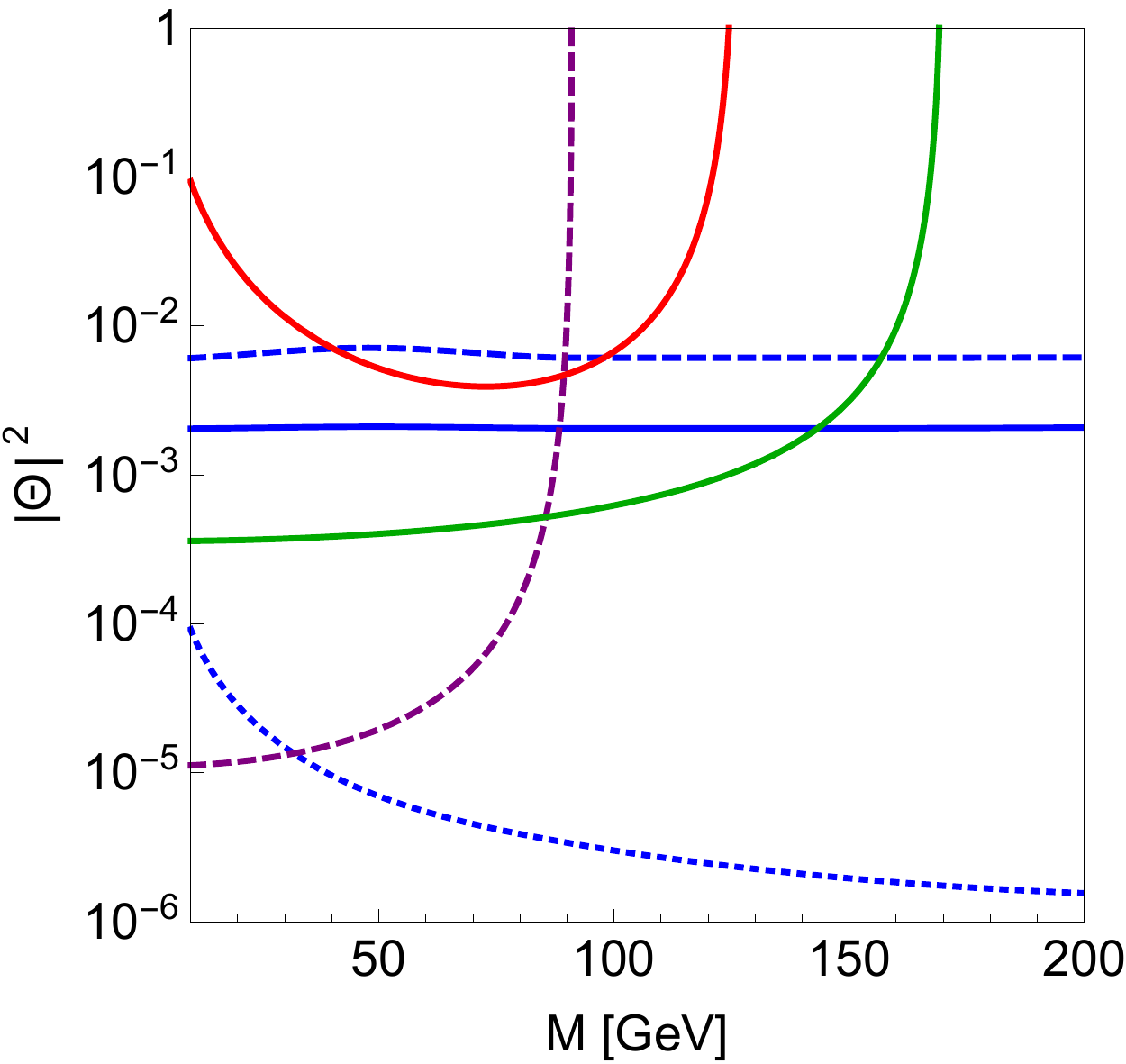}\end{center}
\end{minipage}
\begin{minipage}{0.49\textwidth}
\includegraphics[width=\textwidth]{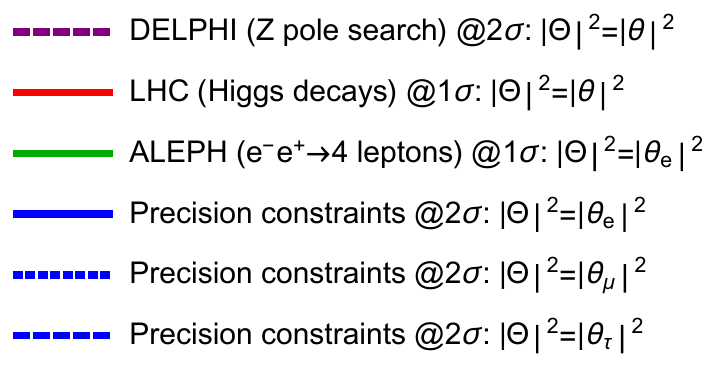}
\end{minipage}
\caption{Present constraints on the active-sterile mixing parameters as a function of the sterile neutrino mass, taken from ref.~\cite{Antusch:2015mia}.}
\label{fig:presentconstraints}
\end{figure}

\noindent
Sterile neutrinos have been and are searched for at past end present colliders, respectively. Below is a list with the most stringent limits from searches at LEP and at the LHC.

\begin{itemize}
\item Search for sterile neutrinos produced in $Z$ boson decays. The strongest bound was derived from the limit on the branching ratio of the $Z$ boson from \cite{Abreu:1996pa}, Opal \cite{Akrawy:1990zq}, Aleph \cite{Decamp:1991uy} and L3 \cite{Adriani:1993gk}. The corresponding upper bound is shown by the purple dashed line in fig.~\ref{fig:presentconstraints}.
\item The measurement of the $W$ boson production cross section allows to limit the production cross section for sterile neutrinos via leptonic final states at LEP-II. Direct searches have also been performed, e.g.\ in ref.\ \cite{Achard:2001qv}. The corresponding upper bound is shown by the solid green line in fig.~\ref{fig:presentconstraints}.
\item The observation of Higgs boson branching ratios at the LHC, in particular of the Higgs to diphoton decays, allow to search for deviations of the Higgs total decay width. The corresponding upper bound is shown by the solid red line in fig.~\ref{fig:presentconstraints}.
\end{itemize}

\section{Heavy neutrino production at $e^-p$ colliders}
\begin{figure}
\begin{minipage}{0.49\textwidth}
\includegraphics[width=0.99\textwidth]{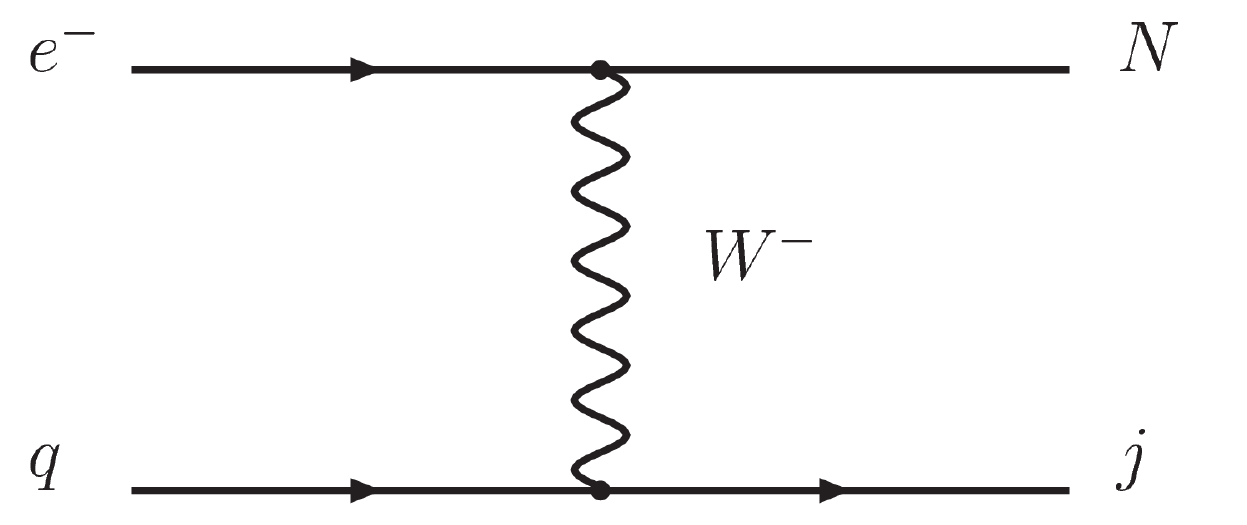}
\end{minipage}
\begin{minipage}{0.49\textwidth}
\includegraphics[width=0.99\textwidth]{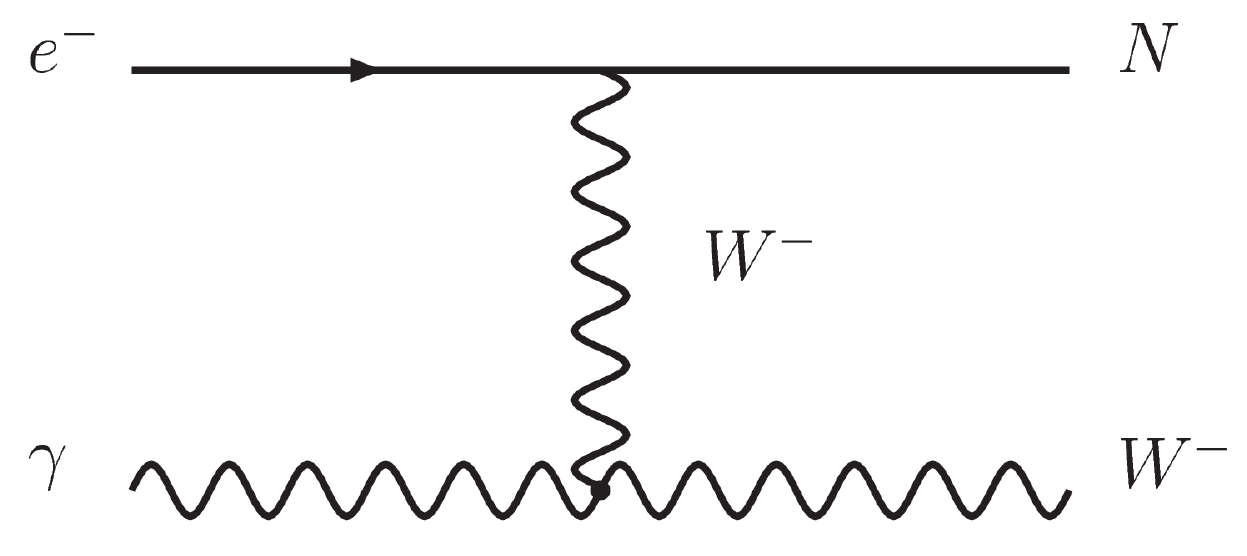}
\end{minipage}
\centering
$\mathbf{W_t^{(q)}}$\hspace{180pt}$\mathbf{W_t^{(\gamma)}}$
\caption{Feynman diagrams representing the leading order production channels for heavy neutrinos in electron-proton scattering.}
\label{fig:Nproduction-ep}
\end{figure}
\noindent
Electron-proton colliders collide protons from a hadron beam with electrons from an intersecting electron beam. They allow for a clean collision environment without pileup at center-of-mass energies of up to a few TeV.

The currently discussed upgrade of the LHC with an electron beam, the Large Hadron-electron Collider (LHeC) \cite{Klein:2009qt,AbelleiraFernandez:2012cc,Bruening:2013bga} is planned to deliver up to 100 fb$^{-1}$ integrated luminosity per year at a center-of-mass energy of $\sim$ 1.0 TeV, collecting $\sim$1~ab$^{-1}$ over its lifetime.
Within the Future Circular Collider design study a more ambitious design is discussed, namely the Future Circular electron-hadron Collider (FCC-eh) \cite{Zimmermann:2014qxa}, which may reuse the a 60~GeV electron beam from the LHeC and could yield center-of-mass energies up to $3.5$~TeV with comparable luminosities to the LHeC, cf.~ref.\ \cite{Klein:2016uwv}.

At electron-proton colliders heavy neutrinos are produced via $t$-channel exchange of a $W$ boson, which we call $\mathbf{W_t}^{(q)}$ (see in fig.~\ref{fig:Nproduction-ep} (left)) production channel, together with a quark jet and we label this channel. 
Another production channel is given by $W\gamma$-fusion, which gives rise to a heavy neutrino and a $W^-$ boson, which we call $\mathbf{W_t}^{(\gamma)}$ (see in fig.~\ref{fig:Nproduction-ep} (right)). The latter is suppressed by the parton distribution function of the photon, but increasingly important for larger center-of-mass energies and sterile neutrino masses.
Both production channels are sensitive on the active-sterile mixing parameter $|\theta_e|$ only.

\begin{figure}
\begin{minipage}{0.49\textwidth}

\includegraphics[width=0.9\textwidth]{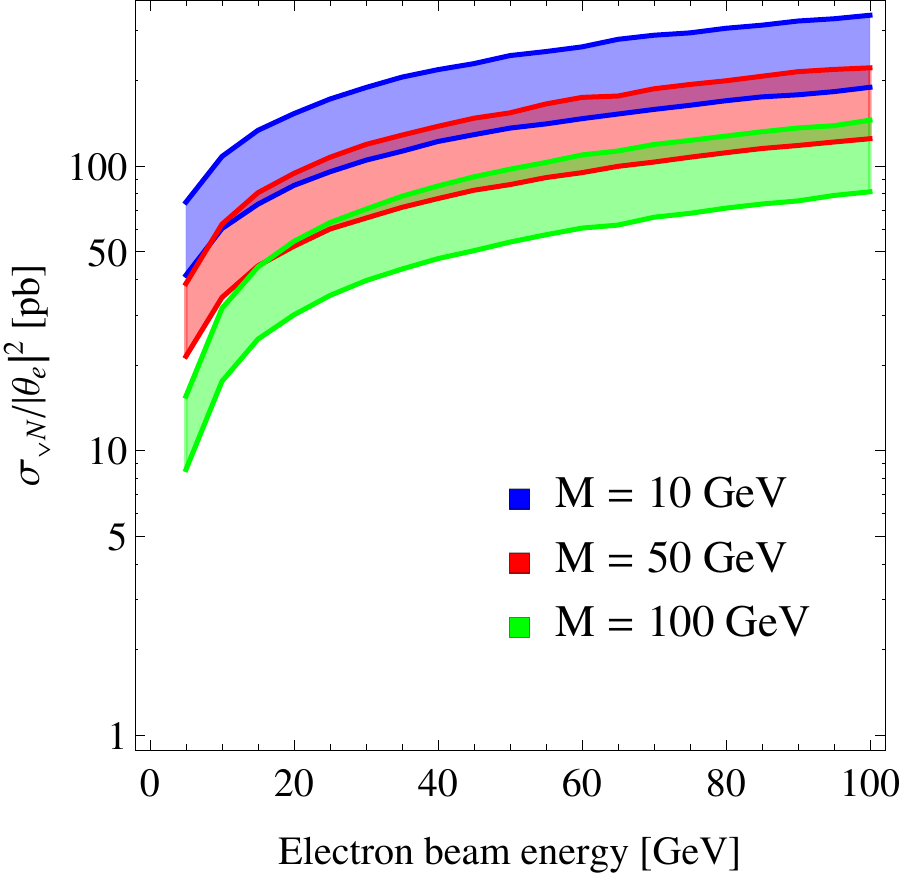}
\begin{center}
{\bf LHeC }
\end{center}
\end{minipage}
\begin{minipage}{0.49\textwidth}

\includegraphics[width=0.9\textwidth]{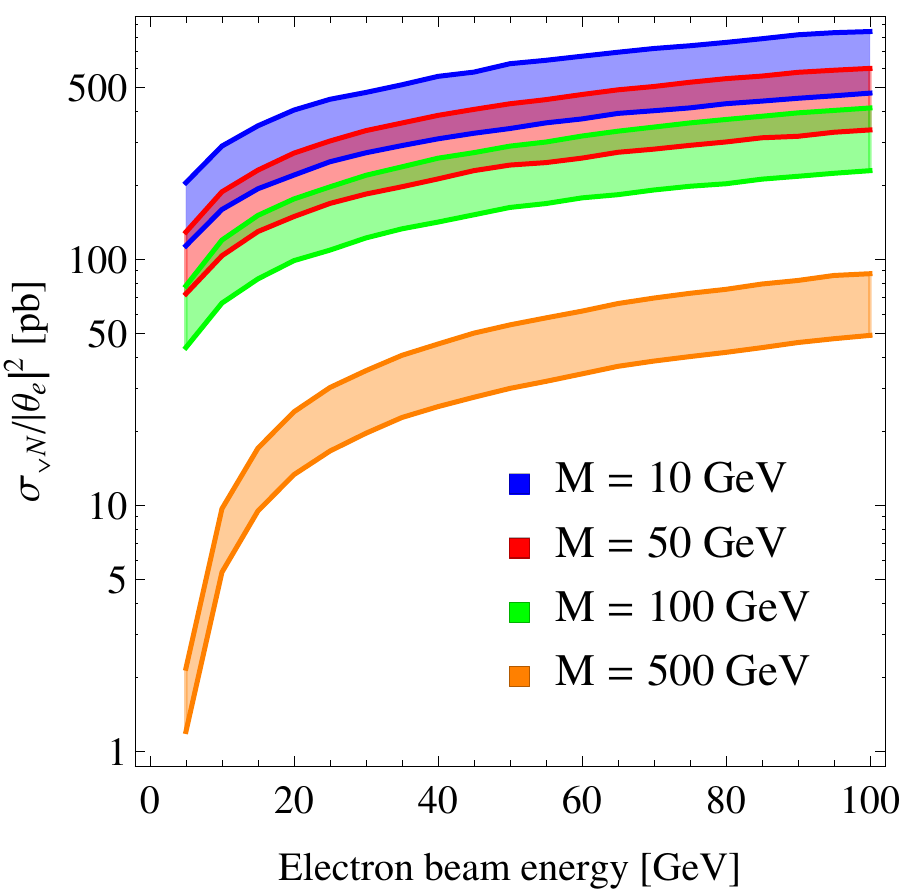}
\begin{center}
{\bf FCC-eh }
\end{center}
\end{minipage}

\caption{Production cross section for the process $\mathbf{W}_t^{(q)}$ of heavy neutrinos at the LHeC (left) and the FCC-eh (right), as a function of the electron beam energy and for different sterile neutrino masses $M$. The width of the colored bands denotes the variation of the electron beam polarisation between 0 and $80\%$.}
\label{fig:xsection-ep}
\end{figure}

We show the production cross section $\sigma_N$ divided by $|\theta_e|^2$ for heavy neutrinos via $\mathbf{W_t}^{(q)}$ in fig.\ \ref{fig:xsection-ep} at the LHeC (left panel) and the FCC-eh (right panel) for different values of the sterile neutrino mass $M$ as a function of the electron beam energy. The width of the colored bands reflects the impact of the beam polarisation, which increases the cross section up to 80\%.

In the following we will consider and use electron beam energies of 60 GeV as benchmark, and integrated luminosities of 1 ab$^{-1}$.

\subsection{Signatures for sterile neutrino searches at $e^-p$ colliders}
\noindent
In the following we provide a summary over the most relevant signatures for sterile neutrinos searches at electron-proton colliders as discussed in detail in ref.~\cite{Antusch:2016ejd}. 

For signatures that stem from the decays of heavy neutrinos that have been produced via the $\mathbf{W_t}^{(q)}$ channel, we generally expect final states with four fermions at the parton level. We note that at the reconstructed level, the exact number of final state objects can vary due to hadronisation of quarks and so-called detector effects. For the sake of clarity we refer to the here discussed signatures as ``four-fermion'' final states and list their signatures in tab.~\ref{tab:signatures4}.

\begin{table}
\begin{center}
\begin{tabular}{|l|c|c|c|}
\hline
Name & Final State & $|\theta_\alpha|$ Dependency & LFV \\ \hline\hline
lepton-trijet  & $jjj \ell_\alpha^-$ & $\dfrac{|\theta_e\theta_\alpha|^2}{\theta^2}$ & $\checkmark$  \\ \hline
jet-dilepton  & $j \ell_\alpha^- \ell_\beta^+ \nu$ &  $\dfrac{|\theta_e\theta_\alpha|^2}{\theta^2}^{(*)}$ & $\checkmark$  \\ \hline
trijet  & $jjj \nu$ & $|\theta_e|^2$ & $\times$  \\ \hline
monojet & $j \nu\nu\nu$ & $|\theta_e|^2$  & $\times$ \\ \hline
\end{tabular}
\end{center}
\caption{Signatures of sterile neutrinos produced via $\mathbf{W}_t^{(q)}$ at $e^-p$ colliders with the corresponding final state, and the dependency on the active-sterile mixing parameters.
A checkmark in the ``LFV'' column indicates that an unambiguous signal for LFV is possible.}
\label{tab:signatures4}
\end{table}

Analogously, the decays of heavy neutrinos that are produced via the $\mathbf{W_t}^{(\gamma)}$ channel gives rise to so-called ``five-fermion'' final states (at the parton level).
We list the corresponding signatures for the ``five-fermion''  final states in tab.~\ref{tab:signatures5}.

\begin{table}
\begin{center}
{\small
\begin{tabular}{|l|c|c|c|}
\hline
Name & Final State & $|\theta_\alpha|$ Dependency & LFV \\ \hline\hline
lepton-quadrijet & $jjjj \ell_\alpha^-$ & $\dfrac{|\theta_e\theta_\alpha|^2}{\theta^2}$  & $\checkmark$ \\ \hline
dilepton-dijet & $\ell_\alpha^-\ell_\beta^+ \nu j j$ & $\dfrac{|\theta_e\theta_\alpha|^2}{\theta^2}^{(*)}$  & $\checkmark$ \\ \hline
trilepton & $\ell_\alpha^- \ell_\beta^- \ell_\gamma^+ \nu\nu$ & $\dfrac{|\theta_e\theta_\alpha|^2}{\theta^2}^{(*)}$  &$\checkmark$  \\ \hline
quadrijet & $jjjj \nu$ & $|\theta_e|^2$  & $\times$ \\ \hline
electron-di-b-jet & $e^- b\bar{b} \nu \nu$ & $|\theta_e|^2$  & $\times$\\ \hline
dijet & $j j \nu\nu\nu$ & $|\theta_e|^2$  & $\times$  \\ \hline
monolepton & $\ell_\alpha^- \nu\nu\nu\nu$ & $|\theta_e|^2$  & $\times$  \\ \hline
\end{tabular}
}
\end{center}
\caption{Signatures of sterile neutrinos produced via $\mathbf{W}_t^{(\gamma)}$ at $e^-p$ colliders with the corresponding final state, and the dependency on the active-sterile mixing parameters.
A checkmark in the ``LFV'' column indicates that an unambiguous signal for LFV is possible.}
\label{tab:signatures5}
\end{table}

Among the signatures listed in tabs.~\ref{tab:signatures4} and \ref{tab:signatures5} the final states without light neutrino, (i.e.\ missing energy on the reconstructed level,) such as the lepton-trijet, allow for unambiguous detection of lepton-flavor violation (LFV). Also the the jet-dilepton final states can give an unambiguous sign of LFV, if the negatively charged lepton is not an electron. Signatures that provide an unambiguous sign for LFV can in principle also provide an unambiguous sign for lepton number violation (LNV), for instance the final state $e^+ jjj$.

As mentioned above, the signatures in tab.\ \ref{tab:signatures5} are more relevant at higher center-of-energies and for larger sterile neutrino masses $M$. They add further possibilities to search for unambiguous signs of LFV (and also LNV).

\subsection{First look: lepton-flavour-conserving signatures}
\begin{figure}
\begin{minipage}{0.52\textwidth}
\includegraphics[width=0.99\textwidth]{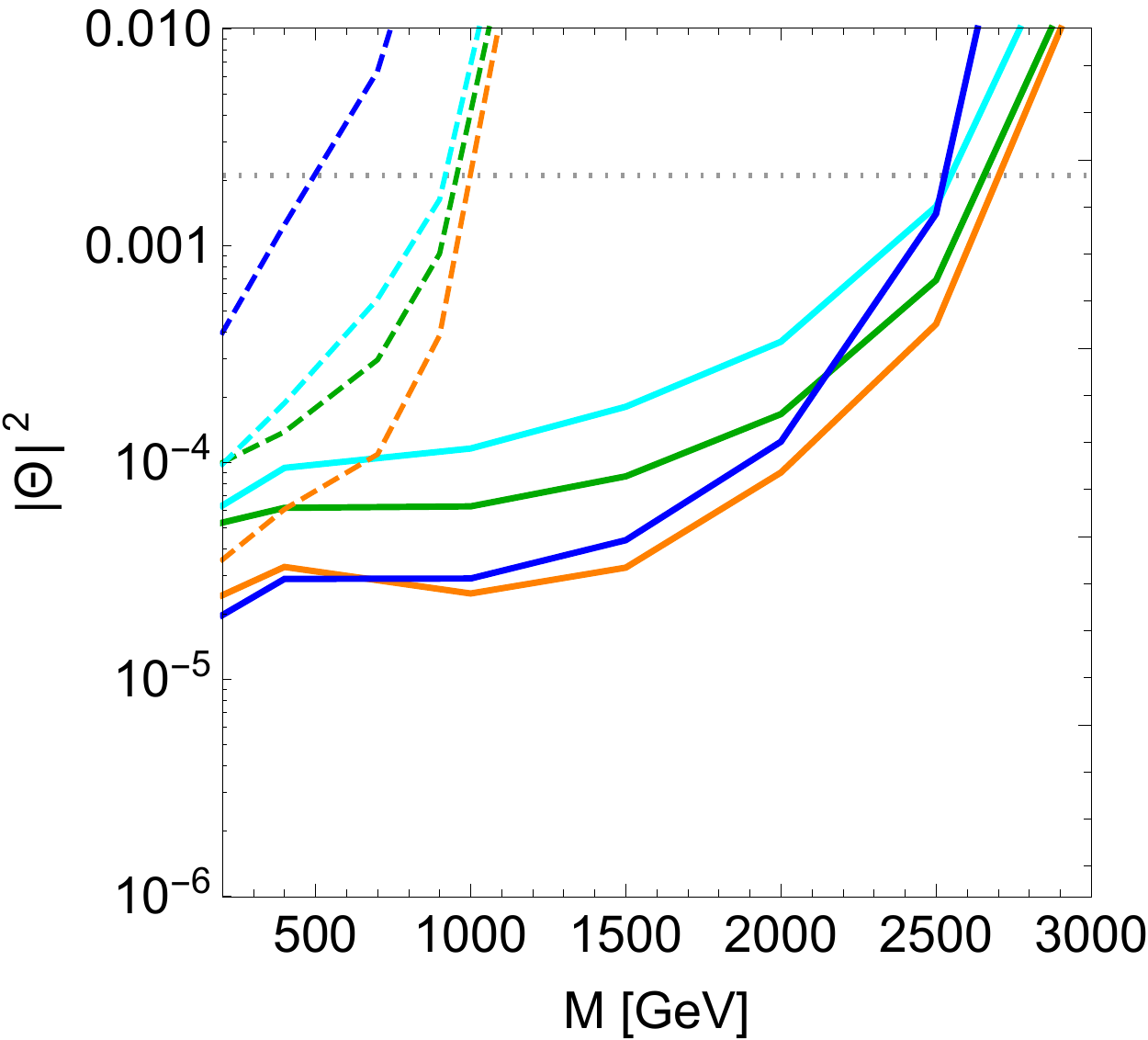}
\end{minipage}
\begin{minipage}{0.46\textwidth}
\includegraphics[width=0.6\textwidth]{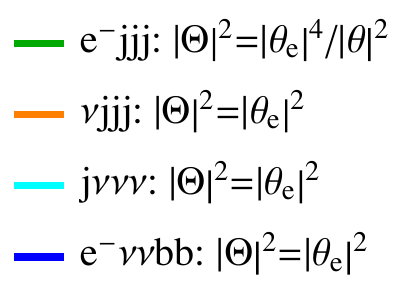}
\end{minipage}
\caption{Estimated sensitivity for sterile neutrino searches via lepton-flavor-conserving signatures at the LHeC (dashed) and the FCC-eh (solid) at $1\sigma$ from ref.~\cite{Antusch:2016ejd}. The blue line labelled $e^- \nu\nu bb$ is produced from the $W\gamma$ fusion channel.}
\label{fig:sensitivity-LFC}
\end{figure}

\noindent
The sensitivities for lepton-number-conserving final states at the LHeC and the FCC-eh have been obtained in ref.~\cite{Antusch:2016ejd} at the parton level. The results from this ``first look'' are presented in fig.\ \ref{fig:sensitivity-LFC} at $1\sigma$  for the lepton-flavor conserving signatures. As mentioned above we assumed an integrated luminosity of 1~ab$^{-1}$ for both, the LHeC (dashed lines) and the FCC-eh (solid lines) in this figure.
In the figure the grey dotted horizontal line denotes the present upper bound on $|\theta_e|$ at the 90\% Bayesian confidence level.

\begin{figure}
\begin{minipage}{0.52\textwidth}
\includegraphics[width=0.99\textwidth]{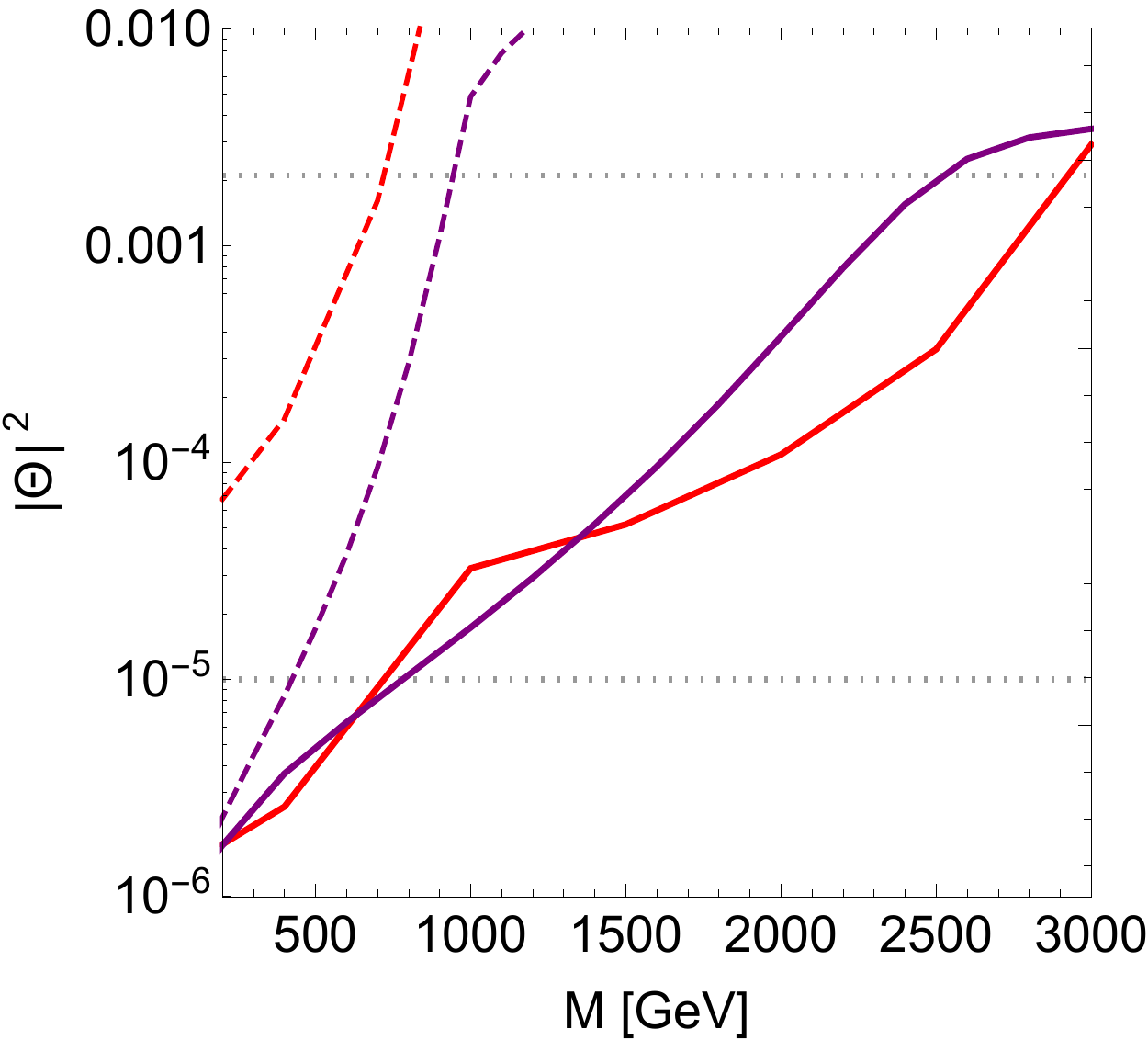}
\end{minipage}
\begin{minipage}{0.46\textwidth}
\includegraphics[width=0.7\textwidth]{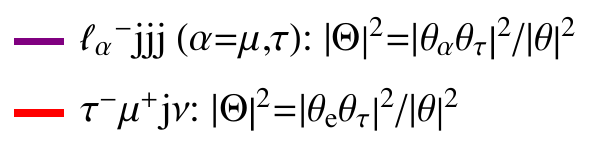}
\end{minipage}
\caption{Estimated sensitivity for sterile neutrino searches via lepton-flavor-violating signatures at the LHeC (dashed) and the FCC-eh (solid) at $1\sigma$ from ref.~\cite{Antusch:2016ejd}.}
\label{fig:sensitivity-LFV}
\end{figure}

The $1 \sigma$ parton level sensitivities for the lepton-flavor violating signatures are presented in fig.\ \ref{fig:sensitivity-LFV}. Therein, the LHeC and the FCC-eh are denoted by dashed and solid lines, respectively. The two black dotted lines denote the present upper bounds on the combinations $|\theta_e \theta_\mu |$ and $|\theta_e \theta_\tau |$ as listed in eq.\ \ref{MUVresult}.

The heavy neutrino production cross section is proportional to $|\theta_e|^2$ and the relative strength of the $|\theta_\alpha|^2$ reflects in the relative strength of the lepton flavors in, e.g.\ the lepton-trijet, and the lepton-quadrijet signature, which both depend on $|\theta_e \theta_\alpha|^2/|\theta|^2$.

The comparison of figs.\ \ref{fig:sensitivity-LFC} and \ref{fig:sensitivity-LFV} shows that the LFV signatures have better expected sensitivities than the LFC ones.
In particular, sterile neutrinos with mixings close to the present upper bound can be tested via the lepton-trijet signature for masses up to $\sim 1$ TeV and $\sim 2.7$ TeV, at the LHeC and FCC-eh, respectively. We note that for $|\theta_\alpha|\sim |\theta_e|$ LFV is something that has to be expected. 

The comparison of the LHeC and FCC-eh shows how the increased proton beam energy improves the sensitivities. Also the relative improvement in sensitivity of the the $W\gamma$ fusion signature $e^- \nu\nu bb$ shows that the higher proton beam energy might bring additional merits by more and competitive signatures for sterile neutrinos searches.

We remark that, although some of the lepton-number-violating signatures do not have SM backgrounds at the parton level, their cross sections are suppressed by the protective symmetry, and we therefore do not expect the resulting sensitivity to be competitive to the LFV signatures.

\subsection{Synergy and Complementarity with other colliders}
\begin{figure}
\centering
\includegraphics[width=0.5\textwidth]{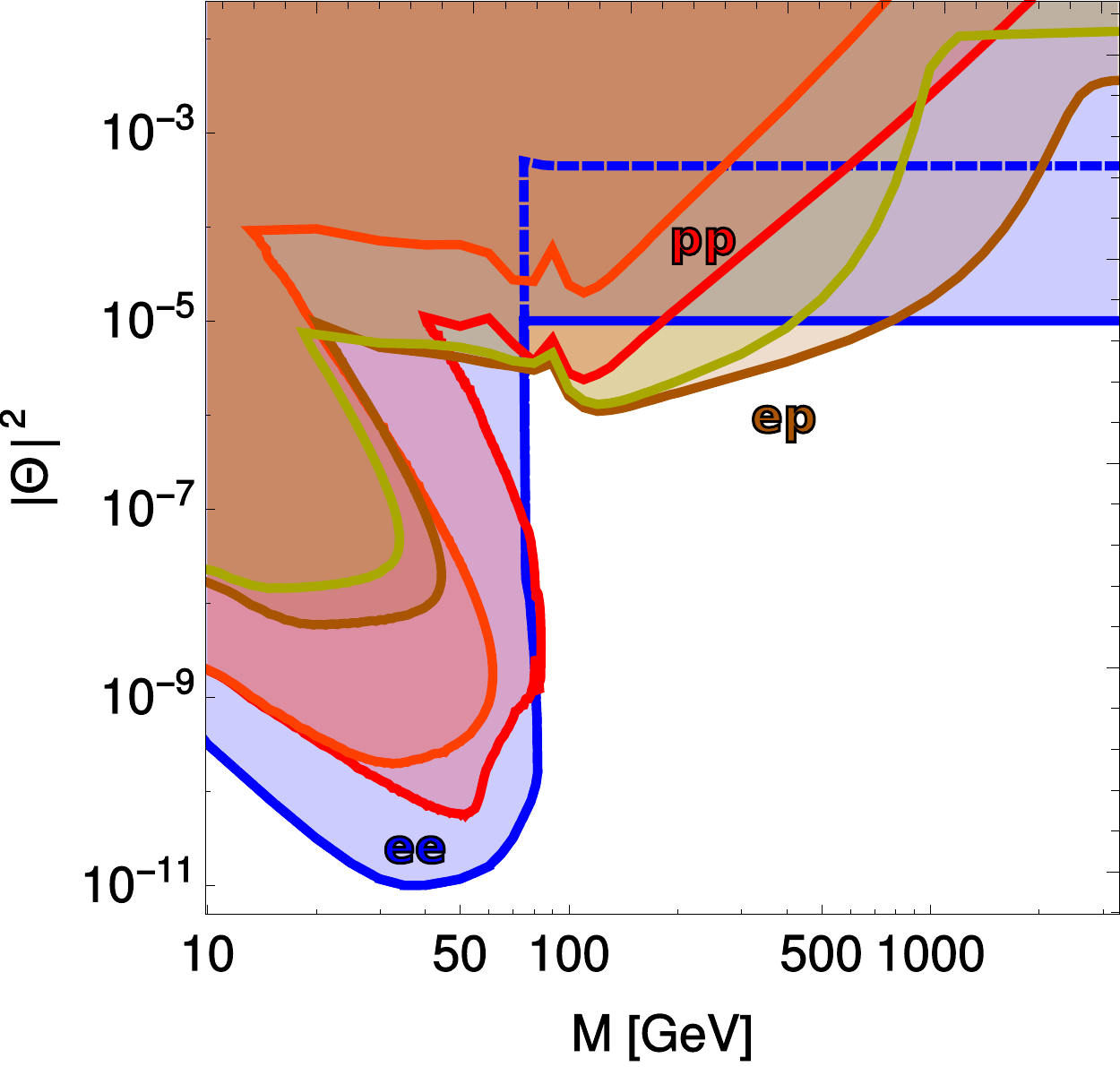}
\caption{Summary of selected estimated sensitivities of the FCC-ee, -hh, and -eh colliders, including the HL-LHC and the LHeC, taken from ref.~\cite{Antusch:2016ejd}.
The FCC-ee sensitivities of the displaced vertex searches and the sensitivity from the EWPO are shown by the blue lines.
Shown in red and dark red are the HL-HLC and the FCC-hh sensitivities, respectively. The estimates for the $e^-p$ colliders, the LHeC in yellow and the FCC-eh in brown have the best prospects for discovering sterile neutrinos via the LFV signatures.}
\label{fig:complementarity}
\end{figure}

\noindent
We show a combination of the different estimated sensitivities for the FCC configurations as well as for the HL-LHC and the LHeC in fig.\ \ref{fig:complementarity}. The sensitivities for the electron-positron collider stem from the displaced vertex searches ($M< m_W$) and from the indirect searches via the EWPO ($M > m_W$) from the Z pole run with 120 ab$^{-1}$, where it is difficult to distinguish possible LNV signatures from LNC ones.
The HL-LHC and the FCC-hh are shown by the red and dark red lines, respectively, and their best sensitivity is expected via the LFV and LNC signature $\ell_\alpha^- \ell_\beta^+ jj$, which is sensitive to different combinations of $|\theta_\alpha \theta_\beta|$ and allows to infer the heavy neutrino mass.
The best sensitivity from direct searches comes from the searches for LFV lepton-trijets at electron-proton colliders, the LHeC and the FCC-eh, shown in yellow and brown, respectively.

\section{Conclusions}
\noindent
Sterile neutrinos are well motivated extensions of the SM, and symmetry protected seesaw scenarios allow for electroweak scale sterile neutrino masses and ${\cal O}(1)$ active-sterile mixings.
Present precision data constrain the active-sterile mixing parameters to $|\theta|^2 \leq {\cal O}(10^{-3})$.

At electron-proton colliders sterile neutrinos are produced dominantly via the mixing with the electron flavor, $|\theta_e|$.
For masses below $m_W$, the best sensitivity can generally be achieved via searches for displaced vertices, which yield sensitivities that are, however, not as good as those at the FCC-ee.
The best sensitivity for masses above $m_W$ is expected to come from the lepton-flavor-violating lepton-trijet final state, which could test mixings down to $|\theta_e \theta_\alpha| \sim 10^{-6}$ for both, the LHeC and the FCC-eh.

We emphasize that the sensitivity from direct searches via the lepton-flavor violating signatures at electron-proton colliders significantly enhances the sensitivity of the HL-LHC and the FCC-hh; it can reach smaller values for the active-sterile mixing parameter $|\theta_e|$ or larger values for the masses.
In particular, the direct searches for sterile neutrinos can tests sterile neutrinos with masses up to 1.0 and 2.7 TeV at the LHeC and the FCC-eh, respectively.

It is important to realize that the direct searches for sterile neutrinos with masses above a few hundred GeV have greater prospects at electron-proton colliders compared to the proton-proton colliders, and also compared to circular electron-positron colliders.
In particular, the improvement of the LHC sensitivity to sterile neutrino searches by the LHeC upgrade is huge.
It therefore seems to be a wasted opportunity not to upgrade a proton-proton collider facility with an electron beam.

\subsection*{Acknowledgements}
This work has been supported by the Swiss National Science Foundation and by the ``Fund for promoting young academic talent'' from the University of Basel under the internal reference number DPA2354. O.F. is thankful to the organizers of the DIS17 in Birmingham for the invitation, and to the participants of the conference for the stimulating atmosphere and the many useful discussions.


\begin{thebibliography}{99}

\bibitem{MarcoDrewes}
	Figure kindly provided by Marco Drewes, private communication.

\bibitem{Minkowski:1977sc} 
  P.~Minkowski,
  Phys.\ Lett.\  {\bf 67B}, 421 (1977).


\bibitem{Mohapatra:1979ia} 
  R.~N.~Mohapatra and G.~Senjanovic,
  Phys.\ Rev.\ Lett.\  {\bf 44}, 912 (1980).


\bibitem{Yanagida:1980xy} 
  T.~Yanagida,
  Prog.\ Theor.\ Phys.\  {\bf 64}, 1103 (1980).


\bibitem{GellMann:1980vs} 
  M.~Gell-Mann, P.~Ramond and R.~Slansky,
  Conf.\ Proc.\ C {\bf 790927}, 315 (1979)
  [arXiv:1306.4669 [hep-th]].


\bibitem{Canetti:2012vf} 
  L.~Canetti, M.~Drewes and M.~Shaposhnikov,
  Phys.\ Rev.\ Lett.\  {\bf 110}, no. 6, 061801 (2013)
  [arXiv:1204.3902 [hep-ph]].


\bibitem{Canetti:2012kh} 
  L.~Canetti, M.~Drewes, T.~Frossard and M.~Shaposhnikov,
  Phys.\ Rev.\ D {\bf 87}, 093006 (2013)
  [arXiv:1208.4607 [hep-ph]].


\bibitem{Heurtier:2016iac} 
  L.~Heurtier and D.~Teresi,
  Phys.\ Rev.\ D {\bf 94}, no. 12, 125022 (2016)
  [arXiv:1607.01798 [hep-ph]].


\bibitem{Davidson:2002qv} 
  S.~Davidson and A.~Ibarra,
  Phys.\ Lett.\ B {\bf 535}, 25 (2002)
  [hep-ph/0202239].


\bibitem{Hambye:2003rt} 
  T.~Hambye, Y.~Lin, A.~Notari, M.~Papucci and A.~Strumia,
  Nucl.\ Phys.\ B {\bf 695}, 169 (2004)
  [hep-ph/0312203].


\bibitem{Pilaftsis:1997jf} 
  A.~Pilaftsis,
  Phys.\ Rev.\ D {\bf 56}, 5431 (1997)
  [hep-ph/9707235].


\bibitem{Pilaftsis:2003gt} 
  A.~Pilaftsis and T.~E.~J.~Underwood,
  Nucl.\ Phys.\ B {\bf 692}, 303 (2004)
  [hep-ph/0309342].


\bibitem{Dev:2014laa} 
  P.~S.~Bhupal Dev, P.~Millington, A.~Pilaftsis and D.~Teresi,
  Nucl.\ Phys.\ B {\bf 886}, 569 (2014)
  [arXiv:1404.1003 [hep-ph]].


\bibitem{Hambye:2016sby} 
  T.~Hambye and D.~Teresi,
  Phys.\ Rev.\ Lett.\  {\bf 117}, no. 9, 091801 (2016)
  [arXiv:1606.00017 [hep-ph]].


\bibitem{Drewes:2012ma} 
  M.~Drewes and B.~Garbrecht,
  JHEP {\bf 1303}, 096 (2013)
  [arXiv:1206.5537 [hep-ph]].


\bibitem{Drewes:2016gmt} 
  M.~Drewes, B.~Garbrecht, D.~Gueter and J.~Klaric,
  JHEP {\bf 1612}, 150 (2016)
  [arXiv:1606.06690 [hep-ph]].


\bibitem{Drewes:2016jae} 
  M.~Drewes, B.~Garbrecht, D.~Gueter and J.~Klaric,
  JHEP {\bf 1708}, 018 (2017)
  [arXiv:1609.09069 [hep-ph]].


\bibitem{Drewes:2013gca} 
  M.~Drewes,
  Int.\ J.\ Mod.\ Phys.\ E {\bf 22}, 1330019 (2013)
  [arXiv:1303.6912 [hep-ph]].


\bibitem{Antusch:2016ejd} 
  S.~Antusch, E.~Cazzato and O.~Fischer,
  Int.\ J.\ Mod.\ Phys.\ A {\bf 32}, no. 14, 1750078 (2017)
  [arXiv:1612.02728 [hep-ph]].


\bibitem{Gavela:2009cd} 
  M.~B.~Gavela, T.~Hambye, D.~Hernandez and P.~Hernandez,
  JHEP {\bf 0909}, 038 (2009)
  [arXiv:0906.1461 [hep-ph]].


\bibitem{Antusch:2015mia} 
  S.~Antusch and O.~Fischer,
  JHEP {\bf 1505}, 053 (2015)
  [arXiv:1502.05915 [hep-ph]].


\bibitem{Antusch:2014woa} 
  S.~Antusch and O.~Fischer,
  JHEP {\bf 1410}, 094 (2014)
  [arXiv:1407.6607 [hep-ph]].


\bibitem{Adam:2013mnn} 
  J.~Adam {\it et al.} [MEG Collaboration],
  Phys.\ Rev.\ Lett.\  {\bf 110}, 201801 (2013)
  [arXiv:1303.0754 [hep-ex]].


\bibitem{Bentz:2009yy} 
  W.~Bentz, I.~C.~Cloet, J.~T.~Londergan and A.~W.~Thomas,
  Phys.\ Lett.\ B {\bf 693}, 462 (2010)
  [arXiv:0908.3198 [nucl-th]].


\bibitem{Erler:2004cx} 
  J.~Erler and M.~J.~Ramsey-Musolf,
  Prog.\ Part.\ Nucl.\ Phys.\  {\bf 54}, 351 (2005)
  [hep-ph/0404291].


\bibitem{Kumar:2013yoa} 
  K.~S.~Kumar, S.~Mantry, W.~J.~Marciano and P.~A.~Souder,
  Ann.\ Rev.\ Nucl.\ Part.\ Sci.\  {\bf 63}, 237 (2013)
  [arXiv:1302.6263 [hep-ex]].


\bibitem{Abreu:1996pa} 
  P.~Abreu {\it et al.} [DELPHI Collaboration],
  Z.\ Phys.\ C {\bf 74}, 57 (1997)
  Erratum: [Z.\ Phys.\ C {\bf 75}, 580 (1997)].


\bibitem{Akrawy:1990zq} 
  M.~Z.~Akrawy {\it et al.} [OPAL Collaboration],
  Phys.\ Lett.\ B {\bf 247}, 448 (1990).


\bibitem{Decamp:1991uy} 
  D.~Decamp {\it et al.} [ALEPH Collaboration],
  Phys.\ Rept.\  {\bf 216}, 253 (1992).


\bibitem{Adriani:1993gk} 
  O.~Adriani {\it et al.} [L3 Collaboration],
  Phys.\ Rept.\  {\bf 236}, 1 (1993).


\bibitem{Achard:2001qv} 
  P.~Achard {\it et al.} [L3 Collaboration],
  Phys.\ Lett.\ B {\bf 517}, 67 (2001)
  [hep-ex/0107014].


\bibitem{Klein:2009qt} 
  M.~Klein,
  arXiv:0908.2877 [hep-ex].


\bibitem{AbelleiraFernandez:2012cc} 
  J.~L.~Abelleira Fernandez {\it et al.} [LHeC Study Group],
  J.\ Phys.\ G {\bf 39}, 075001 (2012)
  [arXiv:1206.2913 [physics.acc-ph]].


\bibitem{Bruening:2013bga} 
  O.~Bruening and M.~Klein,
  Mod.\ Phys.\ Lett.\ A {\bf 28}, no. 16, 1330011 (2013)
  [arXiv:1305.2090 [physics.acc-ph]].


\bibitem{Zimmermann:2014qxa} 
  F.~Zimmermann, M.~Benedikt, D.~Schulte and J.~Wenninger,
  IPAC-2014-MOXAA01.


\bibitem{Klein:2016uwv} 
  M.~Klein,
  Annalen Phys.\  {\bf 528}, 138 (2016).


\end{thebibliography}
\end{document}